\documentstyle[12pt]{article}

\def\cite#1{${}^{#1)}$}

\title{
\begin{flushright}
{\normalsize Yaroslavl State University\\
             Preprint YARU-HE-95/02\\
             hep-ph/9505229} \\[5mm]
\end{flushright}
THREE TYPES OF FERMION MIXING AND POSSIBLE MANIFESTATIONS
OF A PATI--SALAM LEPTOQUARK IN THE LOW--ENERGY PROCESSES}

\vspace{10mm}

\author{A.A.~Gvozdev, A.V.~Kuznetsov, N.V.~Mikheev \\
and L.A.~Vassilevskaya\\
{\small\it Division of Theoretical Physics, Department of Physics,}\\
{\small\it Yaroslavl State University, Sovietskaya 14,}\\
{\small\it 150000 Yaroslavl, Russian Federation}\\
presented by A.V.~Kuznetsov}

\date{May 1995}

\begin{document}

\maketitle

\vglue 5mm

\begin{center}
{\it Talk given at the XXXth Rencontres de Moriond \\ Electroweak
Interactions
and Unified Theories \\ Les Arcs, Savoie, France, March 11-18, 1995}
\end{center}

\newpage

\begin{abstract}

I report the recent studies\cite{1,2} on the low--energy
manifestations of a minimal extension of the Standard Model based on the
quark-lepton symmetry $SU(4)_V \otimes SU(2)_L \otimes G_R$ of the
Pati-Salam type. Given this symmetry the third type of mixing
in the interactions ot the $SU(4)_V$ leptoquarks with quarks and
leptons is shown to be required. An additional arbitrariness of the
mixing parameters could
allow, in principle, to decrease noticeably the lower bound on the vector
leptoquark mass originated from the low-energy rare processes.
The decays $\mu \rightarrow e \gamma$, $\mu \rightarrow e \gamma \gamma$,
and $\mu \rightarrow e e \bar e$ via the vector leptoquark are analysed
and the specific hierarchy of the decay probabilities
$\Gamma(\mu \rightarrow e e \bar e) \gg
\Gamma(\mu \rightarrow e \gamma \gamma) \gg
\Gamma(\mu \rightarrow e \gamma)$ is shown to exist.
The upper limits on the
branching ratios at a level of $10^{-18}$
for the $\mu \rightarrow e \gamma \gamma$ decay
and at a level of $10^{-15}$ for the $\mu \rightarrow e e \bar e$ decay
are established.

\end{abstract}

\vglue 5mm

\section{Introduction}

All existing experimental data in particle physics are in good agreement
with the Standard Model predictions. However, the problems exist which
could not be resolved within the SM and it is obviously not a complete
or final theory. It is unquestionable that the SM should be the low-energy
limit of some higher symmetry. The question is what could be this symmetry.
And the main question is what is the mass scale of this symmetry restoration.
A gloomy prospect is the restoration of this higher symmetry at once on
a very high mass scale, the so-called 'gauge desert'. A concept of a
consecutive symmetry restoration is much more attractive. It looks
natural in this case to suppose a correspondence of the hierarchy
of symmetries and the hierarchy of the mass scales of their restoration.
Now we are on the first step of some stairway of symmetries
and we try to guess what could be the next one.
If we consider some well--known higher symmetries from this point of view,
two questions are pertinent. First, isn't the supersymmetry as the
symmetry of bosons and fermions, higher than the symmetry within the
fermion sector, namely, the quark--lepton symmetry\cite{3}, or the
symmetry within the boson sector, namely, the left--right
symmetry\cite{4}? Second, wouldn't the supersymmetry restoration be
connected with a higher mass scale than the others?

We should like to analyse a possibility when the quark-lepton symmetry
is the next step beyond the SM. We take a minimal symmetry of the
Pati-Salam type with the lepton number as the fourth color\cite{3},
$SU(4)_V \otimes SU(2)_L \otimes G_R$. The fermions are combined into the
fundamental representations of the $SU(4)_V$ group, the neutrinos with
the {\it up} quarks and the charged leptons with the {\it down} quarks.
Some attractive features of this symmetry should be pointed out.

i) The renormalizability of the SM demands some quark-lepton symmetry,
namely, the fermions have to be combined into generations for the
cancellation of the triangle anomalies.

ii) The proton decay is absent.

iii) A natural explanation for the quark fractional hypercharge takes
place.

\noindent Really, the 15-th generator of the $SU(4)$ group can be written
in the form  $\qquad \qquad T_{15} = \sqrt{3/8} \, diag \, ( 1/3 \, , \,
1/3 \, , \, 1/3 \, , \, -1 )$.
It is traceless and the values of the left hypercharge $Y_L$ appear to
be placed on the diagonal. Let us call it the vector hypercharge,
$Y_L = Y_V$.

iv) Let us suppose that $G_R \, = \, U(1)_R$. If we take the
well--known values
of the SM hypercharge of the left and right, and $up$ and $down$ quarks and
leptons, then from the equation $Y_{SM} \, = \, Y_V \, + \, Y_R$ the values
of the right
hypercharge $Y_R$ occur to be equal $\pm 1$ for the $up$ and $down$ fermions,
both quarks and leptons.
It is tempting to interpret this fact as the evidence for the
right hypercharge to be actually the doubled third component of the right
isospin. Hence the $G_R$ group is possibly $SU(2)_R$.

The most exotic object of the Pati--Salam type symmetry is the charged and
colored gauge $X$ boson named leptoquark.
Its mass $M_X$ should be the scale of reducing
of $SU(4)_V$ to $SU(3)_c$. The bounds on the vector leptoquark
mass\cite{5} were obtained from the data on the $\pi \rightarrow e \nu$
decay and from the upper limit on $K^0_L \rightarrow \mu e$ decay.
In fact, these estimations
were not comprehensive because the phenomenon of a mixing in the lepton-quark
currents was not considered there. It can be shown that such a mixing
inevitably occurs in the theory.

\section{Three types of fermion mixing}

Three fermion generations are
combined into the \{4,2\} representations of the semi-simple group
$SU(4)_V \otimes SU(2)_L$ of the type

\begin{equation}
\left ( \begin{array}{c} u^c \\ \nu \end{array} \,
\begin{array}{c} d^c \\ \ell \end{array}
\right )_i , \qquad (i=1,2,3) , \label{eq:d}
\end{equation}

\medskip

\noindent where $c$ is the color index.
The mixing in the quark interaction with the $W$ bosons being depicted by
the Cabibbo-Kobayashi-Maskawa matrix is sure to exist in Nature.
If one starts from the diagonal $d, \nu, \ell$ states and the $u$
states mixed by the CKM matrix than at the one--loop level the $d$
states are mixed due to the conversion $d \to W + u (c,t) \to d'$ and
then the $\ell$ states are mixed also, $\ell \to X + d (s,b) \to \ell'$, etc.
Consequently, it is necessary for the renormalizability of the model
to include all kinds of mixing at the tree-level.
In the general case, none of the $\, u,d,\nu,\ell \,$ components is the mass
eigenstate.
Due to the identity of the three representations~(\ref{eq:d})
they always could be regrouped so that one of the components was
diagonalized with respect to mass. If we diagonalize the charged
lepton mass matrix, then the representations~(\ref{eq:d}) can be rewritten
to the form where the $\nu, u, $ and $d$ states are not the mass
eigenstates and are included
into the same representations as the charged leptons $\, \ell$,
$\nu_\ell  =  {\cal K}_{\ell i} \nu_i , \; u_{\ell}  =
{\cal U}_{\ell p} u_p , \; d_{\ell}  =  {\cal D}_{\ell n} d_n$ .
Here  $\nu_i, u_p, $ and $d_n \, (i, p, n = 1,2,3)$  are the mass
eigenstates, and ${\cal K}_{\ell i} , \, {\cal U}_{\ell p}$, and
${\cal D}_{\ell n}$ are the unitary mixing matrices.
The standard Cabibbo-Kobayashi-Maskawa
matrix is seen to be $V \, = \, {\cal U}^+ \cal D$.
This is as far as we know about $\cal U$ and $\cal D$ matrices.
$\cal K$ is the mixing matrix in the lepton sector.

\section{Bounds from the low--energy experiments}

Subsequent to the spontaneous $SU(4)_V$ symmetry breaking up to $SU(3)_c$
on the $M_X$ scale six massive vector bosons are separated from the 15-plet
of the gauge fields to generate three charged and colored leptoquarks.
Their interaction with the fermions has the form

\begin{equation}
{\cal L}_X \, = \, \frac{g_S(M_X)}{\sqrt 2} \big [
{\cal D}_{\ell n}
\big ( \bar \ell \gamma_{\alpha} d^c_n \big ) +
\big ( {\cal K^+ \; \cal U} \big )_{i p}
\big ( \bar{\nu_i} \gamma_{\alpha} u^c_p \big ) \big ] X^c_{\alpha} +
h.c.
\label{eq:Lx}
\end{equation}

\noindent The constant
\, $g_S(M_X)$ \, can be expressed in terms of the strong coupling constant
\, $\alpha_S$ \, at the leptoquark mass scale
$M_X, \quad g_S^2(M_X)/4 \pi = \alpha_S(M_X)$.

If the momentum transferred is \, $q \ll M_X$, \, then the Lagrangian
{}~(\ref{eq:Lx}) in the second order leads to the effective four-fermion
vector-vector interaction of quarks and leptons. By using the Fiertz
transformation, the scalar,
pseudoscalar, vector and axial-vector terms may be separated in the
effective Lagrangian.
The QCD correction amounts to the appearance of the magnifying factor
$Q(\mu)$ at the scalar and pseudoscalar terms,
$Q(\mu) \, = \, ( {\alpha_S(\mu)}/{\alpha_S(M_X)})^{4/\bar b}$.
Here $\alpha_S(\mu)$ is the effective strong coupling constant
at the hadron mass scale $\mu \sim 1~GeV$,
$\; \bar b \, = \, 11 \, - \, \frac {2}{3} \bar n_f, \; \bar n_f $
is the averaged number of the quark
flavors at the scales $\mu^2 \le q^2 \le M_X^2$. If the condition
$M_X^2 \gg m_t^2$ is valid, then we have $\, \bar n_f \, \simeq \, 6$,
and $\bar b \, \simeq \, 7$.

As the analysis shows, the tightest restrictions on the leptoquark mass
$M_X$ and the mixing matrix $\cal D$ elements
can be obtained from the experimental data on rare $\pi$ and $K$ decays and
$\mu^- \rightarrow e^-$ conversion in nuclei.
In the description of the interactions of $\pi$ and $K$ mesons it is
sufficient to take the scalar and pseudoscalar terms only. As we
shall see later, these terms acquire, in addition to the QCD corrections,
an extra enhancement at the amplitude by the small quark current masses.

One can easily see that the leptoquark contribution to the $\pi
\rightarrow e \nu$ decay is not suppressed by the electron mass
in contrast to the $W-$contribution.
Taking into account the interference of the leptoquark and $W-$
exchange amplitudes we get the following expression for the ratio

\begin{equation}
R \, = \,
\frac{\Gamma (\pi \rightarrow e \nu)}
{\Gamma (\pi \rightarrow \mu \nu)} \, = \,
R_{SM} \left [ 1 \, - \,
\frac{2 \sqrt 2 \pi \, \alpha_S(M_X) \, m^2_{\pi} \, Q(\mu)}
{G_F M^2_X m_e (m_u(\mu) + m_d(\mu) )} \; Re \left ( \frac{{\cal D}_{e d}
{\cal U}^*_{e u}}{V_{u d}} \right ) \right ],
\label{eq:R}
\end{equation}

\noindent where $R_{SM} = (1.2345 \pm 0.0010) \cdot 10^{-4}$ is
the value of the ratio
in the Standard Model\cite{6}, $m_{u,d}(\mu)$ are the running current
masses. To the $\mu \simeq 1~GeV$ scale there correspond the well-known
values $m_u \simeq 4~MeV, m_d \simeq 7~MeV$ and $m_s \simeq 150~MeV$.
Using the experimental data\cite{7},  $R^{exp} = (1.2310 \pm 0.0037)
\cdot 10^{-4}$, we get
the following lower bound on the leptoquark mass

\begin{equation}
M_X > (210~TeV) \cdot |Re  ( {{\cal D}_{e d}{\cal U}^*_{e u}} /
{V_{u d}} ) |^{1/2}.
\label{eq:X1}
\end{equation}

{}From the data on the $K \to e \nu$ decay we obtain similarly

\begin{equation}
M_X > (55~TeV) \cdot |Re ({{\cal D}_{e s}{\cal U}^*_{e u}}
/ {V_{u s}} ) |^{1/2}.
\label{eq:55T}
\end{equation}

One can establish the following limits from the data\cite{8}
on the rare decays $K^0_L \rightarrow \mu e$ and
$K^0_L \rightarrow e^+ e^-$

\begin{equation}
M_X > (1200~TeV) \cdot |{\cal D}_{e d}
{\cal D}^*_{\mu s} \; + \;
{\cal D}_{e s}
{\cal D}^*_{\mu d} |^{1/2}, \qquad
\label{eq:X2}
\end{equation}

\begin{equation}
M_X > (1400~TeV) \cdot |Re \big ( {\cal D}_{e d}
{\cal D}^*_{e s} \big )|^{1/2}.
\label{eq:Kee}
\end{equation}

The situation with another rare $K$ decay,
$K^0_L \rightarrow \mu^+ \mu^-$, is rather intriguing.
The recent measurements of the branching ratio at BNL\cite{9}
lowered its value closely to the unitary limit $Br_{abs} = 6.8 \cdot 10^{-9}$,
and thus the decay amplitude has no real part. However, it was
shown\cite{10} that the real part could not be small in the
SM with a heavy top quark. Isn't it a signal for a new physics, e.g.
leptoquark? In this regard the discontinuance of the experiment KEK E137
where the $K^0_L \rightarrow \mu^+ \mu^-$ decay rate was also measured,
is regrettable.

A low-energy process under an intensive experimental searches, where the
leptoquark could manifest itself is the $\mu e$ conversion in nuclei.
We estimate the branching ratio of the conversion in titanium and
establish the bound on the model parameters
on the base of the experimental data\cite{11}

\begin{equation}
M_X > (680~TeV) \cdot |{\cal D}_{e d}
{\cal D}^*_{\mu d} |^{1/2}.
\label{eq:Xmue}
\end{equation}

The above restrictions on the model parameters contain the elements of the
unknown unitary mixing matrices $\cal D$ and $\cal U$, which are connected
by the condition ${\cal U}^+ {\cal D} = V$ only.
Thus the possibility is not excluded, in principle, that the bounds obtained
did not restrict $\, M_X \,$ at all, e.g. if the elements ${\cal D}_{e d}$
and ${\cal D}_{\mu d}$ were rather small. It would correspond to the
connection of the $\tau$ lepton largely with the $d$ quark in the ${\cal D}$
matrix, and the electron and the muon with the $s$ and $b$ quarks.
In general, it is not contradictory to anything even if it appears to be
unusual. In this case a leptoquark could give a more noticeable
contribution to the flavor-changing decays of the $\tau$ lepton and
$B$ mesons. However, an accuracy of these data is relatively poor.
{}From the experimental limits\cite{12} on the decays
$\tau^- \to \mu^- K^0$,
$\tau^- \to e^- K^0$, and
$B^+ \to K^+ \mu^+ e^-$,
$B^+ \to K^+ \mu^- e^+$,
which are possible via the leptoquark exchange without suppression by the
elements ${\cal D}_{e d}$ and ${\cal D}_{\mu d}$, we obtain

\begin{equation}
M_X > (1~TeV) \cdot |{\cal D}_{\mu s}
{\cal D}^*_{\tau d} |^{1/2},
\qquad
M_X > (1~TeV) \cdot |{\cal D}_{e s}
{\cal D}^*_{\tau d} |^{1/2},
\label{eq:tau}
\end{equation}

\begin{equation}
M_X > (2.4~TeV) \cdot |{\cal D}_{e s}
{\cal D}^*_{\mu b} |^{1/2},
\qquad
M_X > (2.4~TeV) \cdot |{\cal D}_{\mu s}
{\cal D}^*_{e b} |^{1/2}.
\label{eq:BK}
\end{equation}

In the recent paper\cite{13} the limits on the Pati--Salam leptoquark
were also considered in the specific cases when every charged
lepton is connected with only one quark in the currents. For the most part
the results of ref.\cite{13} agree with ours\cite{1}.

\section {Mixing--independent bound}

We could find only one occasion when the mixing-independent lower bound
on the leptoquark mass arises, namely, from the decay
$\pi^0 \rightarrow \nu \bar \nu$. The best laboratory limit\cite{14}
 on this decay is
$\; Br(\pi^0 \rightarrow \nu \bar \nu ) < 8.3 \cdot 10^{-7}$.
In the papers\cite{15} the almost coinciding cosmological and
astrophysical estimations of the width of this decay were found:
$Br(\pi^0 \rightarrow \nu \bar \nu ) < 3 \cdot 10^{-13}$ .
Within the Standard Model this value is proportional to $m^2_{\nu}$.
The process is also possible through the leptoquark mediation, without the
suppression by the smallness of neutrino mass.
On summation over all neutrino species the decay probability
is mixing-independent. As a result the bound on the leptoquark mass occurs
$M_X > 18~TeV$. However, in the recent paper\cite{16} a criticism
has been expressed on both the cosmological and astrophysical limits.
Therefore, only the laboratory limit\cite{14} is reliable to establish
the bound $M_X > 440~GeV$.

\section {Rare muon decays}

The lepton--number violating decays $\mu \rightarrow e \gamma$,
$\mu \rightarrow e \gamma \gamma$, $\mu \rightarrow e e \bar e$
are under the intensive experimental searches. Let us point out,
however, that these decay modes are strongly suppressed in the SM with
lepton mixing due to
the well--known GIM cancellation\cite{17} by the factor
$(m_{\nu}/m_W)^4 \; \sim \; 10^{-39} \cdot
(m_{\nu}/20 \,eV)^4$.

These processes arise in the model with vector leptoquark at the loop
level via the virtual $d, s, b$ quarks\cite{2}.
As the analyses of the radiative muon decays show, the two--photon decay
dominates the one--photon decay

\begin{equation}
\frac{\Gamma(\mu\to e\gamma\gamma)}{\Gamma(\mu\to e\gamma)} \; \sim \;
\frac{\alpha}{\pi} \, \left( \frac{M_X}{m_b} \right) ^{4} \; \gg \; 1.
\label{eq:GIM1}
\end{equation}

\noindent The magnitude of the $\mu \rightarrow e \gamma \gamma$ decay width
could be estimated using the bound~(\ref{eq:Xmue}):

\begin{equation}
Br(\mu \rightarrow e \gamma \gamma) \; < \; 1.0 \cdot 10^{-18}.
\label{eq:BrSD}
\end{equation}

\noindent A similar analysis of the $\mu \to e e \bar e$ decay
within the above restrictions on the model parameters provides

\begin{equation}
Br(\mu \rightarrow e e \bar e)  \; < \; 1.0 \cdot 10^{-15}.
\label{eq:Br3e}
\end{equation}

Although the predicted values of the branches of the $\mu \to e \gamma
\gamma$ and $\mu \rightarrow e e \bar e$ decays
are essentially less then the existing
experimental limits\cite{18}
$Br(\mu \rightarrow e \gamma \gamma)_{exp} < 7.2 \cdot 10^{-11}$,
$Br(\mu \rightarrow e e \bar e)_{exp} < 1.0 \cdot 10^{-12}$,
they are not as small as the predictions of the SM with lepton mixing,
and a hope for their observation in the future still remains.

\section {Conclusions}

$\bullet$ The bounds on the Pati--Salam leptoquark mass were reexamined
with taking account of the mixing in the quark--lepton currents.

\noindent $\bullet$
Some semileptonic meson decays strongly suppressed within the SM
could be induced by vector leptoquark. Their further experimental
investigations are very important.

\noindent $\bullet$
The only mixing independent bound on the vector leptoquark mass arises
from the limits on the invisible $\pi^0 \rightarrow \nu \bar \nu$ decay.
It is $M_X > 440~GeV$ from the laboratory limit and
$M_X > 18~TeV$ from the cosmological limit, but the last has to be verified.

\noindent $\bullet$
The specific hierarchy of the rare muon decay probabilities via vector
leptoquark takes place and the branching ratios are not as small as the
predictions of the SM with lepton mixing.

\bigskip

\noindent {\bf Acknowledgements.}
We are grateful to L.B.~Okun, J.~Ritchie, V.A.~Rubakov, A.D.~Smirnov,
K.A.~Ter-Martirosian and S.~Willenbrock for helpful discussions.

The research described in this publication was made possible in part by
Grant No. RO3000 from the International Science Foundation.

\end{document}